\documentclass[conference, letterpaper]{IEEEtran}
\IEEEoverridecommandlockouts
\usepackage{amsmath,amssymb,amsfonts}
\usepackage{algorithm}
\usepackage{algorithmic}
\usepackage{graphicx}
\usepackage{textcomp}
\usepackage{xcolor}

\usepackage{subcaption}
\usepackage{amsmath,amsthm,amssymb}
\usepackage{floatflt}  
\usepackage[hyphens]{url}

\usepackage[left=1.62cm,right=1.62cm,top=1.8cm]{geometry}

\usepackage{comment}
\usepackage{flushend}

\begin{document}

\title{Securing Health Data on the Blockchain: A Differential Privacy and Federated Learning Framework
}

\author{
\IEEEauthorblockN{1\textsuperscript{st} Daniel Commey}
\IEEEauthorblockA{\textit{Multidisciplinary Engineering} \\
\textit{Texas A\&M University}\\
Texas, USA\\
dcommey@tamu.edu
}
\and
\IEEEauthorblockN{2\textsuperscript{nd} Sena Hounsinou}
\IEEEauthorblockA{\textit{Computer Science \& Cybersecurity} \\
\textit{Metro State University}\\
Minnesota, USA\\
sena.houeto@metrostate.edu
}
\and
\IEEEauthorblockN{3\textsuperscript{rd} Garth V. Crosby}
\IEEEauthorblockA{\textit{Engineering Technology \& Industrial Distribution} \\
\textit{Texas A\&M University}\\
Texas, USA\\
gvcrosby@tamu.edu
}
}

\maketitle

\begin{abstract}
This study proposes a framework to enhance privacy in Blockchain-based Internet of Things (BIoT) systems used in the healthcare sector. The framework addresses the challenge of leveraging health data for analytics while protecting patient privacy. To achieve this, the study integrates Differential Privacy (DP) with Federated Learning (FL) to protect sensitive health data collected by IoT nodes. The proposed framework utilizes dynamic personalization and adaptive noise distribution strategies to balance privacy and data utility. Additionally, blockchain technology ensures secure and transparent aggregation and storage of model updates. Experimental results on the SVHN dataset demonstrate that the proposed framework achieves strong privacy guarantees against various attack scenarios while maintaining high accuracy in health analytics tasks. For 15 rounds of federated learning with an epsilon value of 8.0, the model obtains an accuracy of 64.50\%. The blockchain integration, utilizing Ethereum, Ganache, Web3.py, and IPFS, exhibits an average transaction latency of around 6 seconds and consistent gas consumption across rounds, validating the practicality and feasibility of the proposed approach.
\end{abstract}


\section{Introduction}
\label{sec:introduction}

The rapid advancement of artificial intelligence (AI) has led to significant progress in health analytics, enabling the development of powerful tools for personalized care, disease prediction, and improved health outcomes. The Internet of Things (IoT) has further revolutionized the healthcare industry by facilitating the collection and analysis of vast amounts of health-related data through connected devices and sensors \cite{islam_internet_2015, dimitrov_medical_2016}. However, the sensitive nature of health data raises significant privacy concerns, as the aggregation and analysis of this data pose risks of unauthorized access and breaches \cite{hathaliya_exhaustive_2020}. 

To address these privacy challenges, Federated Learning (FL) has emerged as a promising distributed machine learning framework that allows collaborative model training across multiple devices or servers while keeping the training data localized \cite{mcmahan_communication-efficient_2017, yang_federated_2019}. This approach inherently enhances privacy by reducing the need to share or transmit raw data across the network. However, FL is not immune to privacy threats, such as inference and linkage attacks, where malicious actors could reconstruct individual data points from the shared model updates \cite{melis_exploiting_2019, nasr_comprehensive_2019}.

To further mitigate these privacy vulnerabilities, Differential Privacy (DP) has been proposed as a robust technique to provide formal privacy guarantees by introducing controlled noise to the data or model parameters \cite{dwork_algorithmic_2014, abadi_deep_2016}. Nevertheless, the indiscriminate application of noise can significantly degrade the model's utility, making it challenging to effectively balance privacy and accuracy \cite{jayaraman_evaluating_2019}.

Recent advancements in personalized FL, such as dynamic personalization and adaptive noise distribution strategies, have shown promise in addressing the challenges of data heterogeneity and privacy preservation by adaptively determining the local models' personalized and shared components and applying different noise levels to enhance model performance and robustness \cite{yang_dynamic_2023}. However, these approaches still face comprehensive privacy protection, security, and data integrity limitations.

To further enhance the security and privacy of health data in IoT-based systems, blockchain technology has emerged as a promising solution for secure and transparent data management, giving rise to Blockchain-based Internet of Things (BIoT) systems \cite{fernandez-carames_review_2018, mcghin_blockchain_2019}. Despite these advancements, challenges remain in effectively integrating FL, DP, and blockchain technologies to provide comprehensive solutions that address privacy, security, and data integrity aspects while balancing utility and privacy.

In this work, we propose a framework for health analytics on BIoT data, leveraging the strengths of Federated Learning, Differential Privacy, and blockchain technology to safeguard user privacy while maintaining the utility of the analytics. Our model integrates dynamic personalization and adaptive noise distribution strategies to balance privacy and utility. Furthermore, we employ blockchain technology to ensure secure and transparent aggregation and storage of model updates, enhancing the overall security and integrity of the framework. Integrating blockchain technology adds an extra layer of security and trust, making our framework well-suited for privacy-sensitive applications in the healthcare domain.

Through experiments, we demonstrate that our framework provides strong privacy guarantees against various attack scenarios while retaining high accuracy in health analytics tasks.

The main contributions of this paper are as follows:
\begin{itemize}
\item We propose a framework integrating Federated Learning, Differential Privacy, and blockchain technology for secure and privacy-preserving health analytics on BIoT data.
\item We introduce dynamic personalization and adaptive noise distribution strategies to enhance the flexibility and robustness of our framework in balancing privacy and utility.
\item We leverage blockchain technology to ensure secure and transparent aggregation and storage of model updates, strengthening the security and integrity of the framework.
\item We conduct experiments to demonstrate our proposed framework's strong privacy guarantees and high accuracy under various attack scenarios.
\end{itemize}

The rest of the paper is structured as follows: Section~\ref{sec:related_works} reviews related works, highlighting significant advancements and identifying gaps in the current research landscape. Section~\ref{sec:preliminaries} introduces essential concepts, setting the research framework. In Section~\ref{sec:proposed_model}, we present our innovative framework, focusing on architecture, dynamic personalization, adaptive noise distribution, and blockchain integration. Section~\ref{sec:experiments} evaluates our model's performance, demonstrating its ability to safeguard privacy while preserving data utility. Finally, Section~\ref{sec:conclusion} concludes the paper, summarizing key results and outlining future research avenues.

\section{Related Works}
\label{sec:related_works}

The integration of FL, DP, and BIoT systems has gained significant attention in recent years, particularly in the context of privacy-preserving data analysis and healthcare applications. This section reviews the relevant literature and identifies gaps in the current research landscape.

FL has been explored as a promising approach for collaborative model training while keeping the training data localized, enhancing privacy by reducing the need to share raw data \cite{mcmahan_communication-efficient_2017, yang_federated_2019}. Numerous studies have investigated the application of FL in healthcare, aiming to leverage distributed datasets without compromising patient privacy. Xu et al. \cite{xu_federated_2021} demonstrated the use of FL to improve predictive models for diagnosing diseases from decentralized medical records, while Li et al. \cite{li_multi-site_2020} proposed a privacy-preserving FL framework for medical image analysis.

To further enhance privacy, DP has been proposed to provide formal privacy guarantees by introducing controlled noise to the data or model parameters \cite{dwork_algorithmic_2014, abadi_deep_2016}. Dwork et al. \cite{dwork_differential_2006} introduced the concept of DP and proposed the Laplace mechanism for achieving DP in data analysis. Gu et al. \cite{gu_pckv_2020} proposed a locally differentially private key-value data collection framework that utilizes correlated perturbations to enhance utility.

Blockchain technology has been utilized to ensure the integrity and transparency of data transactions in various sectors, including healthcare. Azaria et al. \cite{azaria_medrec_2016} proposed MedRec, a decentralized record management system for electronic health records using blockchain technology, while Xia et al. \cite{xia_bbds_2017} introduced a blockchain-based data sharing (BBDS) framework for medical data, ensuring secure and transparent data sharing among multiple parties.

The integration of FL, DP, and blockchain technologies offers a comprehensive approach to secure data analytics. Kim et al. \cite{kim_blockchained_2019} proposed a blockchained FL model for IoT networks, ensuring data privacy and security in distributed learning environments.

However, existing works often focus on addressing individual aspects of privacy, security, or data integrity, leaving room for more comprehensive solutions. The dynamic nature of healthcare data and the need for adaptive privacy mechanisms that balance utility and privacy while ensuring integrity have not been fully explored in the context of FL and BIoT systems.

Our proposed framework builds upon the existing literature by introducing a novel integration of FL, DP, and blockchain technologies tailored for healthcare analytics on BIoT systems. We address the specific challenges of privacy, security, and data integrity in a holistic manner, proposing dynamic personalization and adaptive noise distribution mechanisms. By demonstrating the feasibility and effectiveness of our model through experiments, we contribute to the ongoing research efforts in developing secure and privacy-preserving solutions for healthcare data analytics.

\section{Preliminaries}
\label{sec:preliminaries}

This section introduces the key concepts and mathematical foundations of Federated Learning (FL) and Differential Privacy (DP), which form the basis of our proposed framework for secure health data analytics in Blockchain-based IoT (BIoT) systems.

\subsection{Federated Learning}

Federated Learning is a distributed machine learning paradigm that enables collaborative model training across multiple devices or servers without centralizing the training data \cite{mcmahan_communication-efficient_2017, kairouz_advances_2021}. In FL, each participating client trains a local model on their own dataset and shares only the model updates with a central server, which aggregates the updates to improve the global model.

\subsubsection{Problem Formulation}

Consider a set of $K$ clients, each with their own local dataset $D_k$, where $k \in {1, \ldots, K}$. The goal is to learn a global model $\mathbf{w}$ that minimizes the following objective function:

\begin{equation}
\min_{\mathbf{w}} f(\mathbf{w}) = \sum_{k=1}^{K} \frac{n_k}{n} F_k(\mathbf{w}),
\label{eq:global_objective}
\end{equation}

where $n_k = |D_k|$ is the number of data samples in the $k$-th client's dataset, $n = \sum_{k=1}^{K} n_k$ is the total number of data samples across all clients, and $F_k(\mathbf{w})$ is the local objective function of the $k$-th client, defined as:

\begin{equation}
F_k(\mathbf{w}) = \frac{1}{N_k} \sum_{i \in D_k} \ell(\mathbf{w}, \mathbf{x}_i, y_i),
\label{eq:loc_objective}
\end{equation}

where $\ell(\mathbf{w}, \mathbf{x}_i, y_i)$ is the loss function for the $i$-th data sample $(\mathbf{x}_i, y_i)$ in the $k$-th client's dataset.

\subsubsection{Federated Averaging (FedAvg)}
FedAvg \cite{mcmahan_communication-efficient_2017} is a popular algorithm for training models in the FL setting. The algorithm proceeds in rounds, where in each round $t$, a subset of clients $S_t$ is selected to participate in the training. Each selected client $k \in S_t$ performs local training on their dataset $D_k$ for a fixed number of epochs and updates their local model $\mathbf{w}_k^t$. The clients then send their updated local models to the central server, which aggregates them to update the global model $\mathbf{w}^{t+1}$ as follows:

\begin{equation}
\mathbf{w}^{t+1} = \sum_{k \in S_t} \frac{n_k}{n_{S_t}} \mathbf{w}_k^t,
\end{equation}

where $n_{S_t} = \sum_{k \in S_t} n_k$ is the total number of data samples across the selected clients in round $t$. The updated global model is then distributed back to the clients for the next round of local training.

\subsection{Differential Privacy}

Differential Privacy (DP) is a mathematical framework for quantifying the privacy guarantees provided by an algorithm that operates on sensitive data \cite{dwork_algorithmic_2014, dwork_differential_2006}. DP ensures that the output of the algorithm does not depend significantly on any individual data point, thereby limiting the risk of privacy breaches.

\subsubsection{Definition}

A randomized algorithm $\mathcal{M}$ is said to be $(\epsilon, \delta)$-differentially private if for any two neighboring datasets $D$ and $D'$, which differ in at most one data point, and for any subset of outputs $S \subseteq \text{Range}(\mathcal{M})$, the following holds:

\begin{equation}
\Pr[\mathcal{M}(D) \in S] \leq e^\epsilon \cdot \Pr[\mathcal{M}(D') \in S] + \delta,
\end{equation}

where $\epsilon > 0$ is the privacy budget, which controls the level of privacy protection, and $\delta \in [0, 1]$ is a small constant that allows for a slight probability of privacy violation. A smaller $\epsilon$ provides stronger privacy guarantees, while a larger $\epsilon$ allows for more utility in the output.

\subsubsection{Laplace Mechanism}

The Laplace mechanism \cite{dwork_calibrating_2016} is a commonly used technique for achieving $\epsilon$-DP, which involves adding Laplace noise to the output of a function $f$ applied to a dataset $D$. The noise is calibrated to the sensitivity of the function, denoted as $\Delta f$, which is defined as the maximum change in the output of $f$ when applied to any two neighboring datasets $D$ and $D'$:

\begin{equation}
\Delta f = \max_{D, D'} |f(D) - f(D')|_1.
\end{equation}

To achieve $\epsilon$-DP, the Laplace Mechanism adds noise drawn from the Laplace distribution with scale parameter $b = \Delta f / \epsilon$ to the output of $f$:

\begin{equation}
\mathcal{M}(D) = f(D) + \text{Lap}\left(\frac{\Delta f}{\epsilon}\right),
\end{equation}

where $\text{Lap}(b)$ denotes a random variable drawn from the Laplace distribution with mean 0 and scale parameter $b$.

\subsubsection{Gaussian Mechanism}

The Gaussian Mechanism \cite{dwork_algorithmic_2014} is another technique for achieving $(\epsilon, \delta)$-DP, which involves adding Gaussian noise to the output of a function $f$ applied to a dataset $D$. The noise is calibrated to the $L_2$-sensitivity of the function, denoted as $\Delta_2 f$, which is defined as the maximum $L_2$-norm change in the output of $f$ when applied to any two neighboring datasets $D$ and $D'$:

\begin{equation}
\Delta_2 f = \max_{D, D'} |f(D) - f(D')|_2.
\end{equation}

To achieve $(\epsilon, \delta)$-DP, the Gaussian Mechanism adds noise drawn from the Gaussian distribution with standard deviation $\sigma = \Delta_2 f \sqrt{2 \ln(1.25/\delta)} / \epsilon$ to the output of $f$:

\begin{equation}
\mathcal{M}(D) = f(D) + \mathcal{N}\left(0, \sigma^2\right),
\end{equation}

where $\mathcal{N}(0, \sigma^2)$ denotes a random variable drawn from the Gaussian distribution with mean 0 and variance $\sigma^2$.

\subsection{Blockchain Technology}

Blockchain is a decentralized, immutable ledger technology that enables secure and transparent recording of transactions without the need for a central authority \cite{nakamoto_bitcoin_2008}. Our framework utilizes the Ethereum blockchain platform, which supports smart contracts and allows for developing decentralized applications (DApps) \cite{wood_ethereum_2014}.

\subsubsection{Ethereum and Smart Contracts}

Ethereum is an open-source, blockchain-based distributed computing platform that enables the deployment and execution of smart contracts. Smart contracts are self-executing contracts with the terms of the agreement directly written into code \cite{buterin_next-generation_2014}. They are deployed on the Ethereum blockchain and can be interacted with through transactions.

Our framework uses smart contracts to facilitate the secure aggregation and storage of local model updates. The smart contract code is written in Solidity, a high-level programming language specifically designed for implementing smart contracts on the Ethereum blockchain.

\subsubsection{Truffle Framework}

Truffle is a popular development framework for Ethereum that provides a suite of tools for writing, testing, and deploying smart contracts \cite{noauthor_home_2024}. It simplifies the process of building and managing smart contracts by providing a development environment, testing framework, and deployment pipelines.

We leverage Truffle to compile, test, and deploy our smart contracts on the Ethereum blockchain. Truffle also enables the integration of our framework with other tools and libraries in the Ethereum ecosystem, such as Web3.js, for interacting with the blockchain from the client side.

\subsubsection{Consensus Mechanism}

As of September 2022, Ethereum has transitioned from a proof-of-work (PoW) to a proof-of-stake (PoS) consensus mechanism through the Ethereum 2.0 upgrade, also known as "The Merge" \cite{noauthor_merge_nodate}. In PoS, validators are chosen to create new blocks based on the amount of cryptocurrency they have staked as collateral. This transition aims to improve the scalability, security, and energy efficiency of the Ethereum network \cite{buterin_casper_2019}.

However, for local development and testing purposes, we use the Truffle framework with a simulated Ethereum blockchain, such as Ganache, which still employs a PoW-based consensus mechanism. This allows for compatibility with existing development tools and provides a simulated environment for testing our smart contracts before deploying them to the main Ethereum network.

When deploying our smart contracts to the main Ethereum network, they will be subject to the PoS consensus mechanism, ensuring the security and immutability of the stored model updates.

\subsubsection{Gas and Transaction Costs}

In Ethereum, gas is a unit that measures the computational effort required to execute transactions and smart contract operations. Each transaction and smart contract operation consumes a certain amount of gas, which is paid for by the user initiating the transaction in Ether (ETH), the native cryptocurrency of Ethereum \cite{wood_ethereum_2014}.

The total cost of a transaction ($C_t$) can be calculated as:

\begin{equation}
C_t = G_t \times G_p,
\end{equation}

where $G_t$ is the total gas consumed by the transaction, and $G_p$ is the gas price set by the user, typically measured in Gwei (1 Gwei = $10^{-9}$ ETH).

In our framework, gas consumption and transaction costs are important considerations for evaluating the efficiency and practicality of blockchain integration.

\subsection{Dynamic Personalization and Adaptive Noise Distribution}

In our proposed framework, we adapt the dynamic personalization and adaptive noise distribution strategies introduced in \cite{yang_dynamic_2023} to enhance the flexibility and robustness of the federated learning process.

\subsubsection{Dynamic Personalization}

Dynamic personalization aims to adaptively determine the personalized and shared components of the local models based on the importance of each parameter in capturing the local data characteristics. The layer-wise importance measures are computed using the Fisher information matrix (FIM) defined as:

\begin{equation}
\label{eq:fisher_information}
\mathbf{F}_k = \mathbb{E}_{(\mathbf{x}, y) \sim D_k} \left[\nabla_{\mathbf{w}_k} \log p(y|\mathbf{x}, \mathbf{w}_k) \nabla_{\mathbf{w}_k} \log p(y|\mathbf{x}, \mathbf{w}_k)^\top\right],
\end{equation}

where $p(y|\mathbf{x}, \mathbf{w}_k)$ is the predicted probability of label $y$ given input $\mathbf{x}$ and model parameters $\mathbf{w}_k$. The layer-wise importance measure for the $l$-th layer of client $C_k$ is computed as:

\begin{equation}
\label{eq:importance_measure}
s_{k,l} = \frac{\sum_{i \in \mathcal{I}_l} [\mathbf{F}_k]_{i,i}}{\sum_{i=1}^d [\mathbf{F}_k]_{i,i}},
\end{equation}

where $\mathcal{I}_l$ is the set of indices corresponding to the parameters in the $l$-th layer.

Parameters with higher importance measures are retained as personalized components, while the remaining parameters are replaced by the global parameters received from the blockchain network.

\subsubsection{Adaptive Noise Distribution}

Adaptive noise distribution is employed to enhance the robustness of the model against the clipping operation in DP. The local model updates are divided into personalized and shared parameters based on the importance measures. The noisy local model update for client $C_k$ is computed as:

\begin{equation}
\label{eq:noisy_update}
\tilde{\mathbf{w}}_k = \mathbf{w}_k + [\mathbf{n}_u; \mathbf{n}_v],
\end{equation}

where $\mathbf{n}_u \sim \mathcal{N}(0, \sigma_u^2 \mathbf{I}_{d_u})$ and $\mathbf{n}_v \sim \mathcal{N}(0, \sigma_v^2 \mathbf{I}_{d_v})$ are Gaussian noise vectors with different variances $\sigma_u^2$ and $\sigma_v^2$ for personalized and shared parameters, respectively. The variances are determined based on the privacy budget $\epsilon$ and the clipping threshold $C$. The personalized parameters receive less noise to preserve their local information, while the shared parameters receive more noise to align with the clipping threshold.

These dynamic personalization and adaptive noise distribution strategies allow our framework to effectively balance privacy protection and model utility in the context of federated learning and differential privacy.

\section{Proposed Model}
\label{sec:proposed_model}

In this section, we present our novel framework for integrating Federated Learning (FL), Differential Privacy (DP), and blockchain technology to enable secure and privacy-preserving health data analytics in Blockchain-based IoT (BIoT) systems. Our proposed model leverages FL's distributed learning capabilities, DP's privacy guarantees, and blockchain technology's security and transparency.

\subsection{System Architecture}

\begin{figure}[htbp]
\centering
\includegraphics[width=\linewidth]{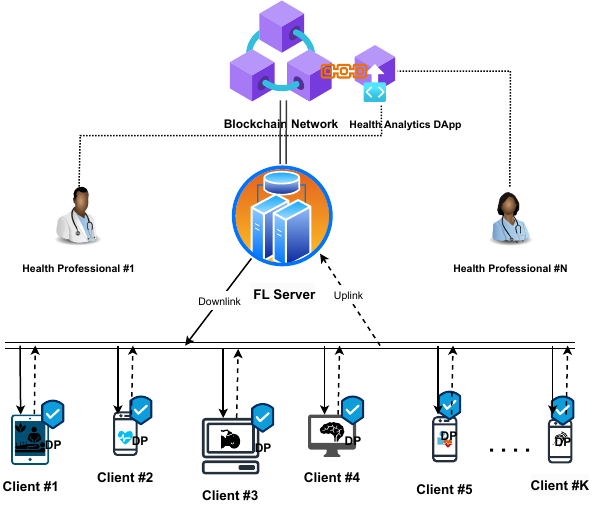}
\caption{System architecture of the proposed model}
\label{fig:architecture}
\end{figure}

The overall architecture of our proposed model consists of three main components: (1) the IoT devices, (2) the FL clients, and (3) the blockchain network. Figure~\ref{fig:architecture} illustrates the system architecture and the interactions between these components.

\subsubsection{IoT Devices}

The IoT devices, such as wearables, smartphones, and medical sensors, collect health-related data from individuals. These devices are responsible for preprocessing the data, involving cleaning, normalization, and feature extraction tasks. The preprocessed data is then securely transmitted to the corresponding FL client associated with the device.

\subsubsection{FL Clients}

The FL clients are the participants in the federated learning process, each holding a local dataset comprised of the preprocessed health data collected by their associated IoT devices. Our model considers a set of $K$ FL clients, denoted as ${C_1, \ldots, C_K}$, where each client $C_k$ has a local dataset $D_k$. The FL clients are responsible for the following tasks:

\begin{itemize}
\item \textbf{Local Model Training}: Each FL client $C_k$ trains a local model $\mathbf{w}_k$ on their local dataset $D_k$ using a predefined learning algorithm (e.g., stochastic gradient descent). The local training process aims to minimize the local objective function $F_k(\mathbf{w}_k)$ in Eq.(\ref{eq:loc_objective}), typically defined as the average loss over the local dataset.

\item \textbf{Dynamic Personalization}: We employ the Fisher information-based approach proposed in \cite{yang_dynamic_2023} to achieve a flexible and adaptive personalization strategy. Each client computes the layer-wise Fisher information of their local model parameters, which reflects the importance of each parameter in capturing the local data characteristics. Parameters with higher Fisher information are retained as personalized components, while the remaining parameters are replaced by the global parameters received from the blockchain network. This dynamic personalization approach ensures that the model adapts to the heterogeneous data distributions across clients while minimizing the impact of DP noise on the personalized parameters.

\item \textbf{Adaptive Noise Distribution}: To enhance the robustness of the model against the clipping operation in DP, we adopt the adaptive noise distribution strategy from \cite{yang_dynamic_2023}. The local model updates are divided into personalized and shared parameters based on the Fisher information. Different levels of Gaussian noise are added to these two sets of parameters, with the personalized parameters receiving less noise to preserve their local information and the shared parameters receiving more noise to align with the clipping threshold. This adaptive noise distribution approach balances privacy protection and model utility.

\item \textbf{Model Update Sharing}: After applying dynamic personalization and adaptive noise distribution, the FL clients share their local model updates with the blockchain network for aggregation. The sharing process is performed securely using cryptographic techniques, such as digital signatures and encryption, to ensure the integrity and confidentiality of the shared updates.

\end{itemize}

\subsubsection{Blockchain Network}

The blockchain network serves as a secure and transparent platform for aggregating the local model updates and maintaining a tamper-proof record of the FL process. The blockchain network consists of a set of nodes, which can be the FL clients themselves or dedicated nodes responsible for consensus and block generation. The main functions of the blockchain network in our model are:

\begin{itemize}
\item \textbf{Model Aggregation}: The blockchain network collects the DP-protected local model updates from the FL clients and performs secure aggregation to obtain the global model update. The aggregation process is typically based on a weighted averaging of the local updates, where the weights are determined by the size of each client's local dataset.

\item \textbf{Model Update Verification}: The blockchain network verifies the integrity and authenticity of the local model updates received from the FL clients. This verification process ensures that legitimate clients generate the updates and have not been tampered with during transmission. Digital signatures and consensus mechanisms, such as proof-of-work or proof-of-stake, are employed to achieve secure verification.

\item \textbf{Model Update Storage}: The verified local model updates and the aggregated global model updates are stored on the blockchain as transactions. Each transaction includes the model update data, timestamp, and cryptographic hash of the previous block, forming an immutable and auditable record of the FL process. The stored model updates can be accessed by authorized parties for verification and analysis purposes.

\end{itemize}

\subsection{Federated Learning Algorithm}

The federated learning process in our proposed model adapts the dynamic personalization and adaptive noise distribution strategies from \cite{yang_dynamic_2023}, with the addition of blockchain integration for secure aggregation and storage. Algorithm \ref{alg:federated_learning} outlines the overall federated learning process.

Before presenting the federated learning algorithm, we introduce our proposed model's mathematical formulation and notations. Let $\mathcal{D} = {D_1, \ldots, D_K}$ denote the set of local datasets held by the $K$ FL clients, where each dataset $D_k = {(\mathbf{x}i, y_i)}{i=1}^{N_k}$ consists of $N_k$ data samples. The goal is to learn a global model $\mathbf{w} \in \mathbb{R}^d$ that minimizes the overall loss function as defined in Equation \ref{eq:global_objective}, where $F_k(\mathbf{w})$ is the local objective function of client $C_k$, as given in Equation \ref{eq:loc_objective}.

In our proposed model, we consider a personalized federated learning setting where each client $C_k$ maintains a local model $\mathbf{w}_k^t = [\mathbf{u}_k^t; \mathbf{v}_k^t]$, consisting of personalized parameters $\mathbf{u}_k^t \in \mathbb{R}^{d_u}$ and shared parameters $\mathbf{v}_k^t \in \mathbb{R}^{d_v}$, with $d = d_u + d_v$. The personalized parameters capture the unique characteristics of the local data, while the shared parameters aim to learn the global patterns across all clients.

\begin{algorithm}[htbp]
\caption{Federated Learning with Dynamic Personalization, Adaptive Noise Distribution, and Blockchain}
\label{alg:federated_learning}
\begin{algorithmic}[1]
\renewcommand{\algorithmicrequire}{\textbf{Input:}}
\renewcommand{\algorithmicensure}{\textbf{Output:}}
\REQUIRE Global epochs $E_g$, local epochs $E_l$, number of clients $K$, privacy budget $\epsilon$, clipping threshold $C$, local datasets $\mathcal{D} = {D_1, \ldots, D_K}$
\ENSURE Global model $\mathbf{w}^{E_g}$
\STATE Initialize global model $\mathbf{w}^0$
\FOR{$t = 1, 2, \ldots, E_g$}
\FOR{each client $C_k$ \textbf{in parallel}}
\STATE Receive global model $\mathbf{w}^{t-1}$ from the blockchain network
\STATE Compute layer-wise importance measures $s_{k,l}$ using Eq. (\ref{eq:fisher_information}) and (\ref{eq:importance_measure})
\STATE Construct personalization masks $\mathbf{M}k$ based on $s{k,l}$ and threshold $\tau$
\STATE Initialize local model $\mathbf{w}_k^t = [\mathbf{M}_k \odot \mathbf{u}_k^{t-1}; (1-\mathbf{M}_k) \odot \mathbf{v}^{t-1}]$
\FOR{$e = 1, 2, \ldots, E_l$}
\STATE Update personalized parameters $\mathbf{u}_k^t$ using $F_k(\mathbf{w}_k^t)$ and regularization
\STATE Update shared parameters $\mathbf{v}_k^t$ using $F_k(\mathbf{w}_k^t)$ and adaptive constraints
\ENDFOR
\STATE Compute local model update $\Delta \mathbf{w}_k^t = \mathbf{w}_k^t - \mathbf{w}_k^{t-1}$
\STATE Clip and add noise to $\Delta \mathbf{w}_k^t$ using Eq. (\ref{eq:noisy_update}) to ensure differential privacy
\ENDFOR
\STATE Aggregate local model updates ${\Delta \tilde{\mathbf{w}}_1^t, \ldots, \Delta \tilde{\mathbf{w}}K^t}$ on the blockchain network
\STATE Update global model $\mathbf{w}^t = \mathbf{w}^{t-1} + \frac{1}{K} \sum{k=1}^K \Delta \tilde{\mathbf{w}}_k^t$
\STATE Store global model update $\mathbf{w}^t$ on the blockchain
\ENDFOR
\RETURN $\mathbf{w}^{E_g}$
\end{algorithmic}
\end{algorithm}

The algorithm begins by initializing the global model $\mathbf{w}^0$ (line 1). In each global epoch $t$, the clients receive the latest global model $\mathbf{w}^{t-1}$ from the blockchain network (line 4). Each client then computes the layer-wise importance measures $s_{k,l}$ using Eq. (\ref{eq:fisher_information}) and (\ref{eq:importance_measure}) (line 5). Personalization masks $\mathbf{M}_k$ are constructed based on the importance measures and a predefined threshold $\tau$ (line 6). The local model $\mathbf{w}_k^t$ is initialized by combining the personalized parameters $\mathbf{u}_k^{t-1}$ masked by $\mathbf{M}_k$ and the shared parameters $\mathbf{v}^{t-1}$ masked by $(1-\mathbf{M}_k)$ (line 7).

During the local training phase, each client performs $E_l$ epochs of training (lines 8-11). The personalized parameters $\mathbf{u}_k^t$ are updated using the local objective function $F_k(\mathbf{w}_k^t)$ and regularization techniques, while the shared parameters $\mathbf{v}_k^t$ are updated using $F_k(\mathbf{w}_k^t)$ and adaptive constraints (lines 9-10).

After local training, each client computes the local model update $\Delta \mathbf{w}_k^t$ (line 12) and applies clipping and noise addition techniques using Eq. (\ref{eq:noisy_update}) to ensure differential privacy (line 13). The noisy local model updates $\Delta \tilde{\mathbf{w}}_k^t$ are then securely aggregated on the blockchain network (line 15), and the global model is updated using the aggregated updates (line 16). The updated global model $\mathbf{w}^t$ is stored on the blockchain for secure record-keeping (line 17).

This process continues for $E_g$ global epochs, and the final global model $\mathbf{w}^{E_g}$ is returned (line 19).

\subsection{Privacy Analysis}

To analyze the privacy guarantees of our proposed model, we consider the following privacy threats:

\begin{itemize}
\item \textbf{Local Data Inference}: An adversary may attempt to infer sensitive information about an individual's health data by observing the local model updates shared by the corresponding FL client. Our model mitigates this threat by applying DP mechanisms to the local model updates before sharing them with the blockchain network. The added noise provides a strong privacy guarantee, ensuring the shared updates do not reveal sensitive information about the individuals' data.

\item \textbf{Membership Inference Attacks}: An adversary may attempt to determine whether a specific individual's data was included in the training dataset of an FL client by observing the model's predictions. Our model counters this threat by leveraging the inherent properties of DP, which guarantees that the model's output does not depend significantly on the presence or absence of any individual data point in the training set.

\end{itemize}

By employing DP mechanisms, secure aggregation techniques, and leveraging the security features of blockchain technology, our proposed model provides strong privacy guarantees against various inference attacks, ensuring the confidentiality of individuals' health data in the FL process.

\subsection{Security Analysis}

The integration of blockchain technology in our model enhances the security and integrity of the FL process. The key security features provided by the blockchain network are:

\begin{itemize}
\item \textbf{Tamper-proof Record}: The blockchain maintains an immutable and auditable record of all local and global model updates throughout the FL process. Each block in the chain is cryptographically linked to the previous block, making it computationally infeasible for an adversary to alter the stored data without detection. This ensures the integrity and traceability of the model updates.

\item \textbf{Consensus Mechanism}: The blockchain network employs a consensus mechanism, such as proof-of-work or proof-of-stake, to validate and approve the addition of new blocks to the chain. The consensus mechanism ensures that all the nodes in the network agree on the state of the blockchain and prevents malicious nodes from tampering with the data or double-spending.

\item \textbf{Cryptographic Security}: The blockchain utilizes cryptographic primitives, such as digital signatures and hash functions, to ensure the data's authenticity and integrity. Digital signatures enable the verification of the origin and ownership of the local model updates. At the same time, hash functions provide a unique and irreversible representation of the block data, making detecting tampering attempts easy.

\end{itemize}

By leveraging blockchain technology's security features, our model ensures the integrity, traceability, and authenticity of the FL process, preventing unauthorized modifications and attacks on shared model updates.

\section{Experimental Evaluation}
\label{sec:experiments}

\subsection{Experimental Setup}
\label{subsec:experimental_setup}

The experiments were conducted on a Lenovo ThinkStation P330 Tiny with the following specifications:

\begin{itemize}
\item Memory: 32 GiB
\item CPU: Intel Core i7-8700T @ 2.40GHz
\item Graphics Card: NVIDIA Quadro P1000
\item Operating System: Ubuntu 22.04.4 LTS
\end{itemize}

We followed the details provided in \cite{yang_dynamic_2023} to implement our proposed framework. We set the learning rate to 1e-3 and optimized the hyperparameters $\tau$, $\lambda_1$, and $\lambda_2$ through grid search in the range of {0.05, 0.1, 0.3, 0.5}. The Rényi Differential Privacy (RDP) algorithm from the Opacus library was employed as the privacy accountant. The value of $\delta$ in ($\epsilon$, $\delta$)-differential privacy was set to the reciprocal of the number of clients (i.e., 1/K).

We conducted experiments on the SVHN dataset \cite{netzer_reading_2011}, a real-world image dataset for digit recognition. Although the SVHN dataset is not directly related to healthcare data, it serves as a suitable benchmark for evaluating the performance of our federated learning framework with personalized differential privacy and blockchain integration. The SVHN dataset has a similar structure to medical image datasets, consisting of labeled images with varying quality and resolution. By demonstrating the effectiveness of our framework on the SVHN dataset, we can infer its applicability to health-related image data, such as X-rays or MRI scans. In future work, we plan to extend our experiments to medical datasets to validate the performance of our framework in the specific context of BIoT systems for healthcare applications.

The experiments were performed using federated learning with personalized differential privacy. We evaluated the impact of different privacy budget values (epsilon) on the model's performance, considering epsilon values of 1.0, 2.0, 4.0, and 8.0. The model was trained for 15 global epochs (communication rounds), with each client performing local training for 3 epochs per round.

We integrated blockchain technology into our federated learning system to enhance data security and integrity. For testing and experimentation purposes, we utilized the Ethereum framework with the Ganache development blockchain. The Web3.py library interacted with the Ethereum blockchain, enabling the deployment and interaction with smart contracts.

We employed the InterPlanetary File System (IPFS) \cite{benet_ipfs_2014}, a decentralized storage solution for large model updates. The model updates were stored on IPFS, and smart contracts recorded the corresponding IPFS hashes on the blockchain. This hybrid approach allows for efficient storage and retrieval of large data while maintaining the security and immutability guarantees the blockchain provides.

The performance metrics evaluated in the experiments include model accuracy, loss, transaction latency, gas consumption, and IPFS data size. These metrics provide insights into the effectiveness of the federated learning process, the impact of differential privacy, and the efficiency of the blockchain integration.

\subsection{Federated Learning with Personalized Differential Privacy}

\subsubsection{Results and Analysis}

\begin{figure}[htbp]
    \centering
    \includegraphics[width=\linewidth]{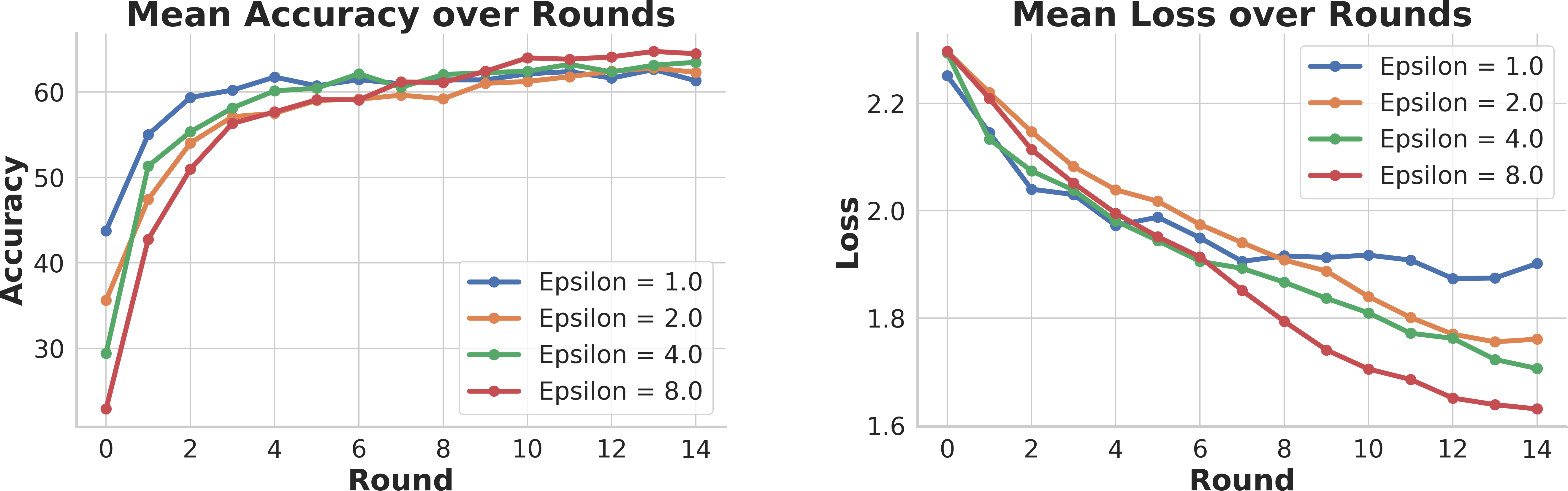}
    \caption{Mean Accuracy and Loss over Rounds for Different Epsilon Values}
    \label{fig:accuracy_loss}
\end{figure}

\paragraph{Model Accuracy}
Figure \ref{fig:accuracy_loss} (left) shows the mean accuracy of the federated learning model over multiple rounds for different epsilon values. As expected, higher epsilon values generally lead to better accuracy, as they allow for more information to be shared during the learning process.

With an epsilon value of 8.0, the model achieves the highest accuracy, starting at around 22.89\% in the first round and gradually increasing to 64.50\% by the final round. This indicates that a larger privacy budget allows for more effective learning and improved model performance.

As the epsilon value decreases, the model's accuracy also decreases. For epsilon values of 4.0, 2.0, and 1.0, the final accuracies are around 63.50\%, 62.31\%, and 61.33\%, respectively. However, it's worth noting that even with a lower epsilon value of 1.0, the model still achieves a reasonable accuracy of over 60\%, demonstrating the effectiveness of federated learning with personalized differential privacy.

\paragraph{Model Loss}
Figure \ref{fig:accuracy_loss} (right) depicts the mean loss of the federated learning model over multiple rounds for different epsilon values. The loss curves show a decreasing trend over the rounds, indicating that the model is learning and improving its performance.

Interestingly, the loss curves for different epsilon values are relatively close, suggesting that the privacy budget has a less significant impact on the model's loss than its impact on accuracy. This observation implies that even with a smaller privacy budget, the model can still effectively minimize the loss and converge to a stable state.

The final loss values for epsilon values of 8.0, 4.0, 2.0, and 1.0 are approximately 1.63, 1.71, 1.76, and 1.90, respectively. While higher epsilon values result in slightly lower loss values, the differences are not as pronounced as in the case of accuracy.

\subsection{Blockchain Experiment and Results}

\subsubsection{Transaction Latency and IPFS Data Size}
\begin{figure}[htbp]
\centering
\includegraphics[width=\linewidth]{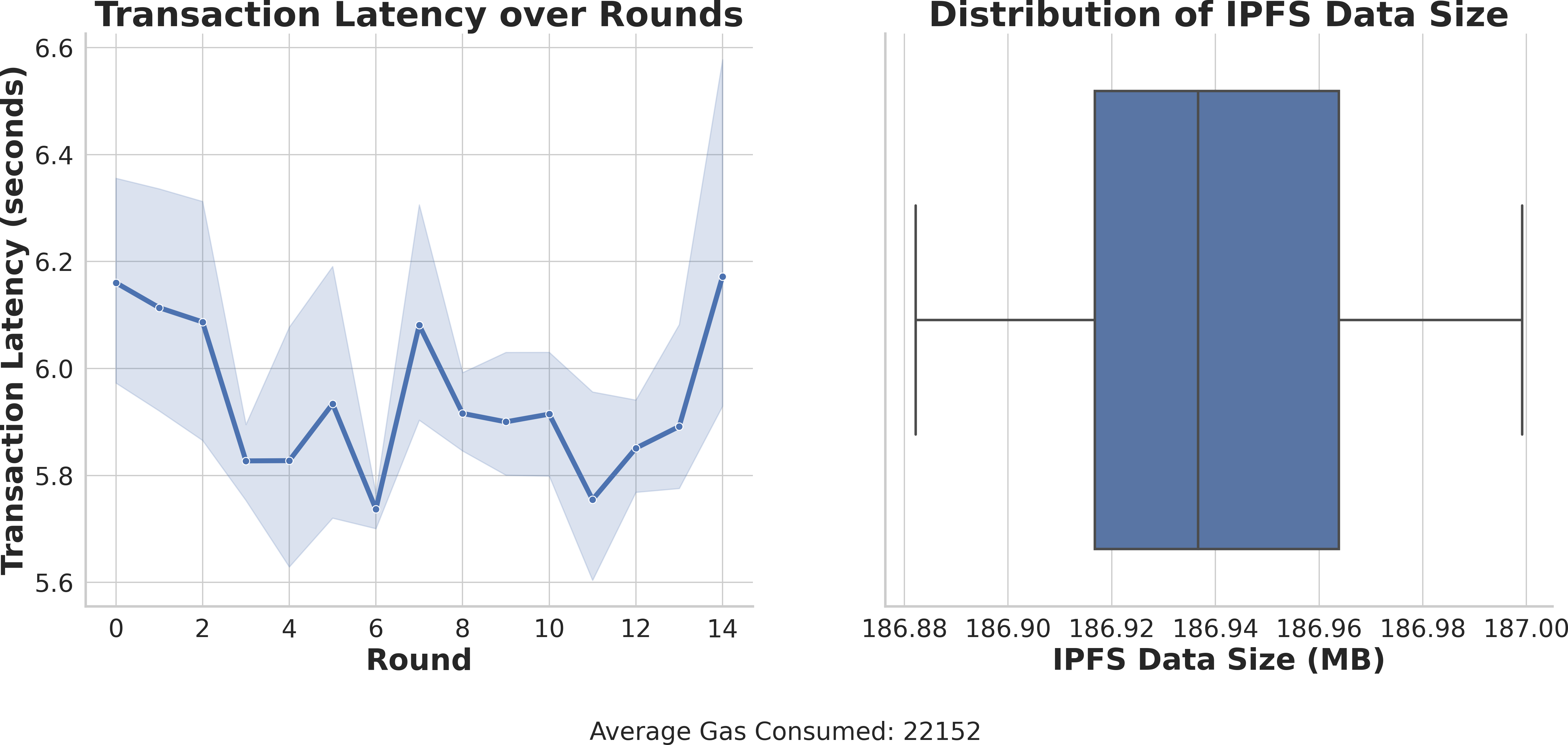}
\caption{Transaction Latency over Rounds and Distribution of IPFS Data Size}
\label{fig:metrics}
\end{figure}

Figure \ref{fig:metrics} presents a combined visualization of the transaction latency over the rounds of federated learning and the distribution of IPFS data sizes. Transaction latency represents the time taken for a transaction (storing a model update) to be confirmed on the blockchain. As can be observed from the plot, the transaction latency exhibits some variation across rounds, with an average latency of around 6 seconds. This latency is acceptable for our federated learning system, considering the security and integrity benefits provided by the blockchain.

The box plot in Figure \ref{fig:metrics} reveals that the IPFS data sizes are concentrated around 196 MB, with minimal variation. This indicates that the model updates have a consistent size throughout the federated learning process. The compact distribution of data sizes demonstrates the efficiency of IPFS in storing the model updates.

\subsubsection{Gas Consumption}
Gas consumption is a measure of the computational resources required to execute transactions on the Ethereum blockchain. In our experiment, we observed that the gas consumption remained constant throughout the rounds, with an average value of 22,152 gas units per transaction. This indicates that the cost of storing model updates on the blockchain is predictable and stable.

The consistent gas consumption across rounds can be attributed to the fixed size of the data being stored on the blockchain. As the model updates have a similar size in each round, the gas required to process the transactions remains the same.

\subsection{Discussion}
The experimental results demonstrate the successful integration of blockchain technology with our federated learning system. By leveraging the Ethereum framework, Ganache, Web3.py, and IPFS, we achieved a secure and transparent mechanism for storing and verifying model updates.

Combining on-chain and off-chain storage, the hybrid storage approach proved an effective solution. IPFS efficiently handled the storage of large model updates while the blockchain maintained the data's immutability and security. The transaction latency and gas consumption metrics indicate the practicality and feasibility of the blockchain integration.

Our main contribution lies in incorporating blockchain technology into the federated learning process. By focusing on the implementation of blockchain, we aim to address the challenges of data security, transparency, and decentralization in health data analytics. The experimental results validate the effectiveness of our approach in achieving these goals.

\section{Conclusion}
\label{sec:conclusion}

In this study, we proposed a novel framework integrating federated learning, differential privacy, and blockchain technology for secure health data analytics in BIoT systems. Our main contribution lies in incorporating blockchain, using Ethereum, Ganache, Web3.py, and IPFS to ensure data security, transparency, and decentralization.

The experimental results on the SVHN dataset demonstrate the effectiveness of our approach in achieving a secure and transparent mechanism for storing and verifying model updates. The hybrid storage solution, combining on-chain and off-chain storage, proved efficient and practical, as evidenced by the transaction latency and gas consumption metrics.

Our work showcases the feasibility and benefits of integrating blockchain technology with federated learning for healthcare applications. The proposed framework addresses the challenges of data privacy, security, and decentralization in BIoT systems, offering a comprehensive solution for secure and collaborative data analytics.

Future research directions include optimizing the scalability and efficiency of blockchain integration, extending the framework to support heterogeneous architectures, and exploring its applicability to diverse healthcare datasets and tasks. Our work lays the foundation for developing privacy-preserving and decentralized, federated learning solutions in healthcare.

\bibliographystyle{IEEEtran}
\bibliography{references}

\end{document}